\documentclass[runningheads]{llncs}
\usepackage{graphicx}
\usepackage{bm}
\usepackage{amsmath} 
\usepackage{amssymb} 
\usepackage{subcaption}
\usepackage{siunitx}

\newcommand{\ie}{i.\,e.~}
\newcommand{\eg}{e.\,g.~}

\newcommand{\Fref}[1]{Fig.~\ref{#1}}
\newcommand{\Eref}[1]{Eq.~\ref{#1}}

\renewcommand{\deg}{^\circ}

\usepackage{color} 
\newcommand{\revision}[1]{{#1}}

\newif\ifanonymized

\begin{document}
%

\anonymizedfalse 

\title{Learning to Avoid Poor Images: Towards Task-aware C-arm Cone-beam CT Trajectories}
\titlerunning{Towards Task-aware C-arm Cone-beam CT Trajectories}
%

\ifanonymized 
\else
\author{Jan-Nico~Zaech\inst{1,2,3}, Cong~Gao\inst{1}, Bastian~Bier\inst{1,2}, Russell~Taylor\inst{1}, Andreas~Maier\inst{2}, Nassir~Navab\inst{1}, and Mathias~Unberath\inst{1}}
\authorrunning{J-N. Zaech et al.}
\fi
%

%
\ifanonymized 
\else
\institute{
Laboratory for Computational Sensing and Robotics, Johns Hopkins University
\and
Pattern Recognition Lab, Friedrich-Alexander-Universit{\"a}t Erlangen-N{\"u}rnberg
\and
Computer Vision Laboratory, Eidgen{\"o}ssische Technische Hochschule Z{\"u}rich
}
\fi
\maketitle              

\begin{abstract}
Metal artifacts in computed tomography (CT) arise from a mismatch between physics of image formation and idealized assumptions during tomographic reconstruction. These artifacts are particularly strong around metal implants, inhibiting widespread adoption of 3D cone-beam CT (CBCT) despite clear opportunity for intra-operative verification of implant positioning, \eg in spinal fusion surgery. 
On synthetic and real data, we demonstrate that much of the artifact can be avoided by acquiring better data for reconstruction in a task-aware and patient-specific manner, and describe the first step towards the envisioned task-aware CBCT protocol. The traditional short-scan CBCT trajectory is planar, with little room for scene-specific adjustment. We extend this trajectory by autonomously adjusting out-of-plane angulation. This enables C-arm source trajectories that are scene-specific in that they avoid acquiring ''poor images'', characterized by beam hardening, photon starvation, and noise. The recommendation of ideal out-of-plane angulation is performed on-the-fly using a deep convolutional neural network that regresses a detectability-rank derived from imaging physics.
\end{abstract}
\section{Introduction}

\paragraph{Background:} Spinal fusion surgery is an operative therapy for chronic back pain with high economic burden~\cite{bmus2014} that is projected to further increase due to our aging society and our increasingly inactive lifestyle. Despite substantial improvements in operative technique, spinal fusion surgery remains high-risk: In addition to usual complications, pedicle screws that breach cortex can result in nerve damage~\cite{gelalis2012accuracy}. Surprisingly, the number of misplaced pedicle screws remains high~\cite{gelalis2012accuracy,cordemans_accuracy_2017}: Cortical breach occurs in  up to 31\% and 72\% of the cases for freehand and fluoroscopy-guided techniques, respectively. Even when surgical navigation is employed, up to 19\% of the screws are not fully contained in cortex~\cite{gelalis2012accuracy}. Currently, screw placement is assessed on post-operative CT images, such that immediate repositioning of implants is not possible. Although intra-operative 3D cone-beam CT (CBCT) imaging using mobile and robotic C-arm X-ray systems is becoming widely available, it is not currently being used for spinal fusion 3D imaging, because compared to CT, C-arm CBCT images suffer from substantially stronger metal artifacts around the highly-attenuating titanium implants, which compromise the value of intra-operative CBCT for assessing cortical breach~\cite{cordemans_accuracy_2017}.

\begin{figure}[tb]
    \centering
    \includegraphics[width=.8\linewidth]{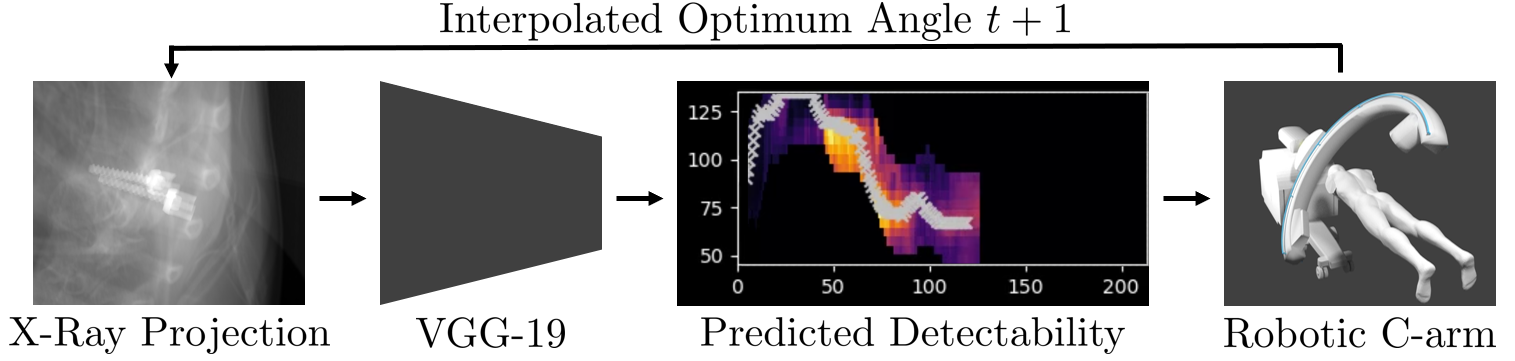}
    \caption{High-level overview of our task-aware trajectory recommendation.}
    \label{fig:pipeline_overview}
\end{figure}

The obvious implication is that image quality must be improved, before CBCT is ready for prime-time in high-volume applications, such as spinal fusion. Most current methods that seek to lift CBCT reconstruction quality to the ''clinical acceptance threshold'' limit themselves to contain artifact propagation (\eg via masking) or image-enhancement (\eg streak reduction)~\cite{gjesteby2016metal}. These methods have in common that they try to deal with artifacts after acquisition of the CBCT short-scan is already completed, and are thus limited by the already corrupted information present in the acquired X-ray projection images. 

This somewhat straight-forward realization implies that there lies huge, unexploited potential in ''simply acquiring better data'' to push the limits of CBCT image quality. In a more formal way, metal artifacts arise from a mismatch between the forward and inverse model, \ie physical effects governing image formation and idealized assumptions made in the tomographic reconstruction algorithm. Sampling data that is less affected by un-explained corruption processes during reconstruction will yield a better conditioned inverse problem, and as an immediate consequence, improved image quality without any additional post-processing. The first step towards the envisioned task- and anatomy-aware CBCT imaging protocol can be realized easily. The traditional short-scan CBCT trajectory~\cite{feldkamp1984practical} is embedded in a single plane, and therefore, provides little room for scene-specific adjustment. We propose to extend this trajectory by autonomously adjusting out-of-plane angulation, which enables C-arm source trajectories that are task-aware and scene-specific in that they avoid acquiring images with substantial corruption (beam hardening and noise) as shown in \Fref{fig:pipeline_overview}. The recommendation and adjustment of ideal out-of-plane angulation is performed on-the-fly using a deep convolutional neural network (ConvNet) that only relies on the current 2D X-ray projection image.

\paragraph{Related Work:} Overall, there is little work on acquisition parameter-side image quality enhancement.  Previous work on task-based trajectories~\cite{stayman_task-based_2013} 
leveraged preoperative CT scans and optimization techniques to select optimal parameters. During application, these approaches would require registration between preoperative CT volume and intra-operative C-arm system, which cannot be achieved easily in practice. Besides this requirement, an even more important limitation is the fact that surgery will alter the patient's anatomy represented by pre-operative CT in an unpredictable way. Therefore, computing task-optimality based on preoperative CT volumes can only serve as a coarse approximation of the true optimal trajectory. These assumptions become even stronger as surgical tools may still be present in the scene~\cite{stayman_task-based_2013}, as tools will strongly affect the optimal solution due to their high attenuation.

The prospect of altering acquisition parameters to improve image quality has recently also been recognized for magnetic resonance imaging~\cite{bahadir2019learning}, where the undersampling pattern in k-space can be optimized via end-to-end learning with respect to fully sampled image. The approach closest to ours considers finding an optimal acoustic window for cardiac ultrasound~\cite{milletari2019straight}, where the current image is interpreted by a reinforcement learning agent that suggests an ultrasound probe displacement towards a better acoustic window. While both previous approaches have some similarity from a conceptual standpoint, they focus on image appearance rather than imaging physics and both the magnetic resonance and ultrasound acquisition protocols are substantially different from CBCT.

\section{Methods}

\begin{figure}[tb]
    \centering
    \includegraphics[width=\linewidth]{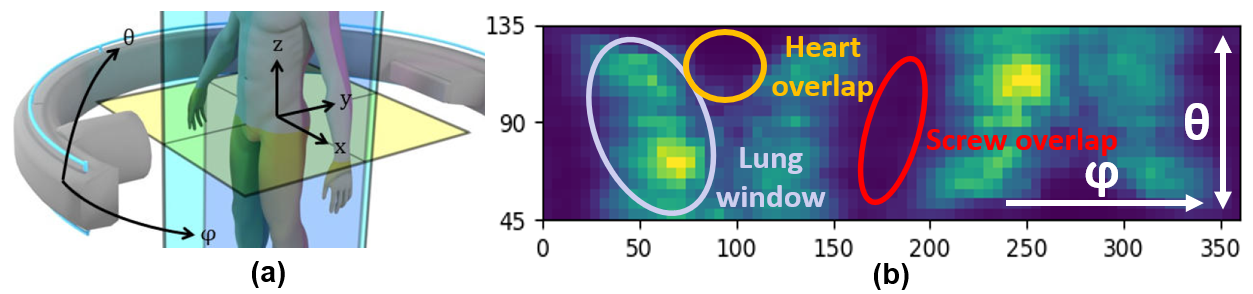}
    \caption{(a) Diminished coordinate system for CBCT: $\varphi$ describes in-plane gantry rotation (traditional short-scan) while $\theta$ is the out-of-plane angle to be adjusted in a task-aware and autonomous fashion. (b) Detectability map for test CT volume as per \Eref{EQ:detectability} (brighter values indicate better detectability). We also show some areas with extremal detectability that can be interpreted semantically.}
    \label{fig:synthetic_detectability_map}
\end{figure}

\paragraph{Assigning Task-optimality Rank to Projection Images:}
Task-based trajectory optimization relies on finding projection images that result in optimum reconstruction quality for a specific task. Therefore, the pipeline is contingent on 1) assigning a task-optimality rank to images in projection domain and 2) selecting views that are optimal in this metric. Previous work used the non-prewhitening matched filter observer derived for penalized likelihood reconstruction~\cite{stayman_task-based_2013} to calculate a detectability index that correlates well with human performance in detection tasks. Calculating detectability requires the patient volume and knowledge about the task, \ie an accurately annotated preoperative CT volume. Using the observer model, the detectability $d^2$ can be calculated as
\begin{equation}\label{EQ:detectability}
d^2(\varphi,\theta) = \frac{\left[\int\int\int |{MTF}(\varphi,\theta)|^2~|W_\text{task}|^2 df_x df_y df_z\right]^2}{\int\int\int {NPS}(\varphi,\theta)~|{MTF}(\varphi,\theta)|^2~|W_\text{task}|^2 df_x df_y df_z},
\end{equation}
where $W_\text{task}$ is the Fourier transform of the region of interest to be imaged with highest quality. Further, ${MTF}$ is the modulation transfer function and ${NPS}$ is the noise power spectrum (we refer to~\cite{gang_task-based_2014} for details) that both depend on the projection image, and therefore, the relative pose of the C-arm with respect to anatomy. Here we consider a diminished C-arm coordinate system $(\varphi,\theta)$ shown in \Fref{fig:synthetic_detectability_map}(a), where $\varphi$ and $\theta$ describe the in- and out-of-plane angle, respectively. It is worth mentioning that in order to compute the above detectability measure for a particular view $(\varphi,\theta)$, the corresponding X-ray projection image must either be simulated from the CT volume or available otherwise.

\paragraph{Predicting Task-Optimal Views from Live Data:} 
If the 3D patient anatomy is perfectly known, the complete trajectory can be optimized for directly. However, this is not the case in surgical environments, where optimal view prediction may only depend on the current and previous 2D X-ray projections. Following recent work in robotics and control, we interpret the detectability index of each possible next view as the quality function and use a ConvNet to directly regress it from the current view. Then, acquiring an optimized trajectory is achieved by selecting the out-of-plane angle with the highest predicted detectability, adjusting $\theta$ as the C-arm gantry moves to the next in-plane angle $\varphi$, and acquiring the next X-ray image at this position, that is then fed back into the ConvNet. Next possible views are defined as views with an increment of $5\deg$ in sweep direction $\phi$. The ConvNet predicts detectability for out-of-plane angles between $\pm25\deg$, uniformly discretized in 11 steps of $5\deg$. This definition allows generating a training dataset, where all meaningful X-ray projections together with their detectability are sampled on a uniform grid with stepsize of $5\deg$ in both $\varphi$ and $\theta$. \revision{The resulting trajectory is patient specific, as the input images used to predict the detectabilities reflect the patient's anatomy at the current point in time.}

Our ConvNet is based on VGG-19~\cite{simonyan_very_2014} with modifications to perform regression instead of classification. Initial weights are pre-trained on ImageNet and the ConvNet is subsequently retrained on our task. \revision{Due to the very short inference time of in the range of few $10^{-2}$\,s on current GPUs, the VGG-19 network is compatible with the near real-time requirements of CBCT acquisition protocols.} During application, angle increments between two views are usually below $1 \deg$, and we use linear interpolation to predict the next best angle. 
The complete pipeline is shown in \Fref{fig:pipeline_overview}, where the image from the {CBCT} system is fed into the ConvNet to predict the detectability of possible next views. Based on this prediction, the interpolation module provides the next out-of-plane angle to the {CBCT} system, that is servoed to the new position, acquires a new X-ray image, and thus, closes the loop.

\paragraph{Training:} Training is performed on realistic digitally reconstructed radiographs (DRRs) generated from CT using the open-source tool DeepDRR~\cite{unberath_deepdrr_2018}. \revision{The pipeline was chosen as it enables the simulation of metal artifact as well as the transfer to real data with only low degradation of prediction performance \cite{Unberath2019}.} For DRR generation, five chest CT-volumes were obtained from the Cancer Imaging Archive. In every volume, six pairs of pedicle screws were annotated and simulated leading to a total of 30 different anatomical configurations, since only a single vertebral level is considered at once. Data augmentation is performed in 3D by randomly varying the C-arm isocenter, yielding a dataset of 212 ''scans'' with 290,016 images in total. A ''scan'' consists of 1368 images uniformly sampled on the truncated sphere with $\varphi \in [0\deg,360\deg)$ and $\theta \in [45\deg,135\deg]$, with the detectability calculated as per \Eref{EQ:detectability} for each image. To guarantee patient independence of training and test, splitting of data is performed on CT level, where four volumes (176 scans) are used for training and one volume (36 scans) is used for test and validation.

The images are saved both noise-free and with noise corresponding to a fluence of 20k\,photons emitted towards every pixel. The noise free images are used to calculate ground-truth detectability while the noisy images are used as input during training of the neural network. This approach was chosen, such that the detectability maps are the optimum learning target, while the network becomes invariant to noisy observations as they would occur in real X-ray projections.

For experiments on real data, a set of analytic phantoms (squares, cylinders, screw model) that represent the chest was implemented and a second \emph{in silico} dataset was generated. \revision{For data generation, dimensions and location of the phantom components where randomly varied within reasonable bounds to reflect the fact that the anatomy, present during inference, is not known at training time.} The dataset consists of 75 ''scans'', generated with the same setup as described for the synthetic data experiments, except for the photon-fluence of 500\,photons per pixel, adapted to the smaller size of the phantom.

\section{Experiments, Results, and Discussion}
\begin{figure}[tb]
    \centering
    \includegraphics[width=.85\linewidth]{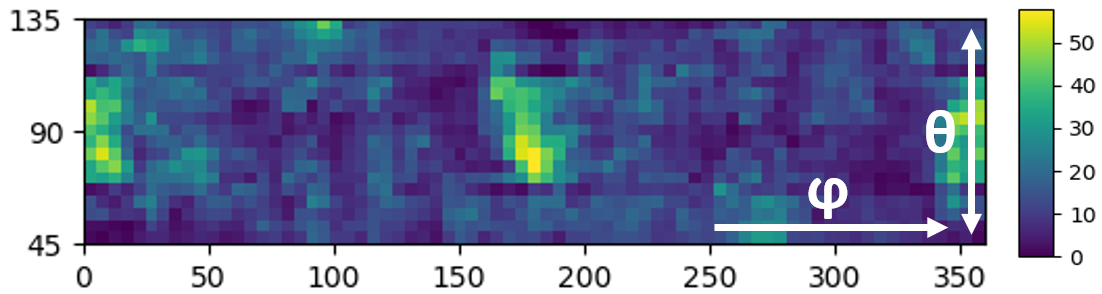}
    \caption{Distribution of degrading detectability in \% for the synthetic test set.}
    \label{fig:spatial_distr_detectability_error}
\end{figure}

\paragraph{Quantitative Synthetic Data Experiments:} Our trajectory optimization pipeline was tested on six different vertebral levels in the separate test volume. As direct evaluation of the training loss function would not represent the quality of the selected trajectory, two surrogate measures were defined. The angular distance of the predicted next-best action to the best action selected from the groundtruth data measures the spatial difference between the predicted and optimal trajectory. While the angular error is an intuitive and interpretable measure, it does not fully capture the performance of the algorithm. Even if the angular distance of the selected action is high, it can still result in a close to optimal reconstruction performance, as the function of detectability values can be multimodal. Therefore, the difference in detectability between the predicted next action and the optimal next action is introduced as a better measure for reconstruction performance. On the test set, \revision{containing 36 scans across 6 different anatomical sites,} these performance metrics evaluated to $8.35\deg\pm11.61\deg$ for the angular distance error and $13.69\%\pm\,18.92\%$ degradation in detectability. The spatial distribution of the average degradation of detectability in the test set is illustrated in \Fref{fig:spatial_distr_detectability_error}.

Besides the quality of selected actions, pipeline stability w.r.t. noise is crucial for practical applications. We compare the average distance between trajectories predicted from noise-free data to noisy samples generated with 400k, 100k, and 50k\,photons per pixel. These measures are $0.83\deg\pm1.56\deg$, $1.13\deg\pm1.63\deg$, and $1.64\deg\pm1.73\deg$ for the different noise levels, respectively.

When the C-arm is used for intra-operative imaging, no optimal alignment between the scanner and the patient's anatomy can be ensured. Therefore, robustness for different initialization angles is a major requirement, \ie the algorithm should transition into the same or equivalent trajectory irrespective of its initialization. Theoretically this can be approached via the Markov property of the proposed prediction pipeline: The detectabilities used for optimizing the trajectory only depend on the last acquired X-ray projection, not on the history of the trajectory. Therefore, as soon as two trajectories would intersect each other, they will merge into a single trajectory.

\begin{figure}[tb]
\centering
\begin{subfigure}[t]{0.24\textwidth}
\centering
\includegraphics[trim= 50 100 150 100,clip,width=\textwidth]{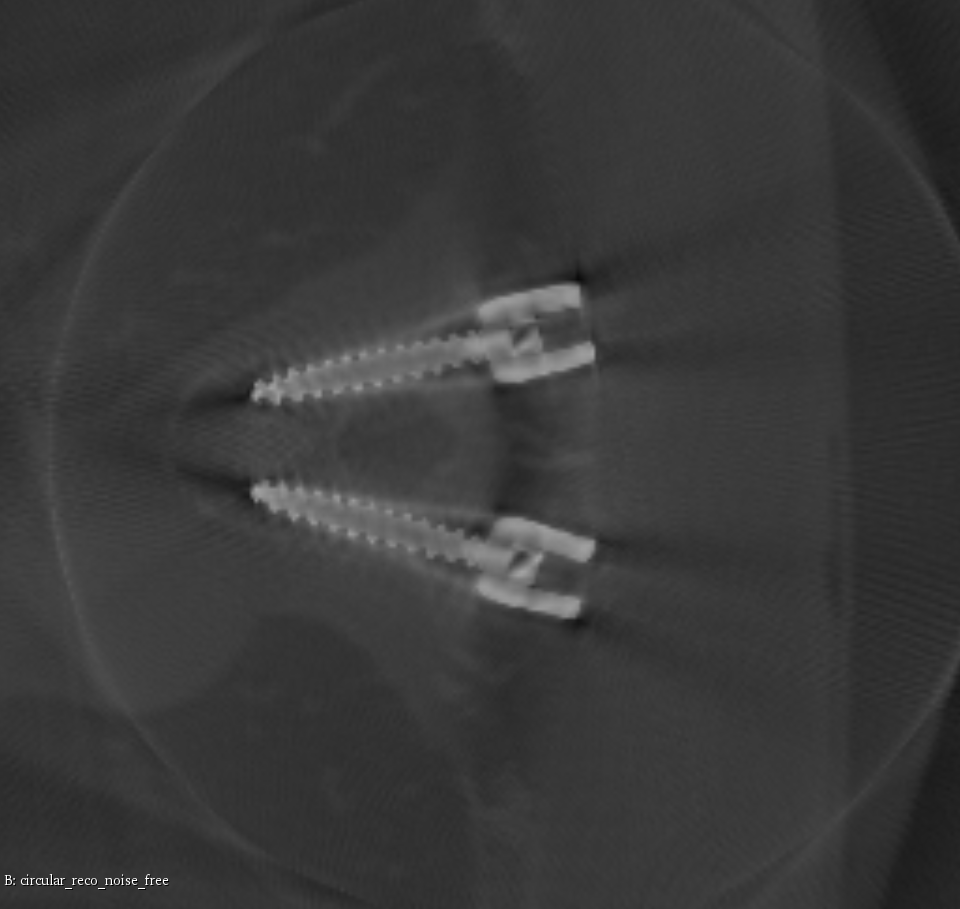}
\caption{}
\end{subfigure}
\begin{subfigure}[t]{0.24\textwidth}
\centering
\includegraphics[trim= 50 100 150 100,clip,width=\textwidth]{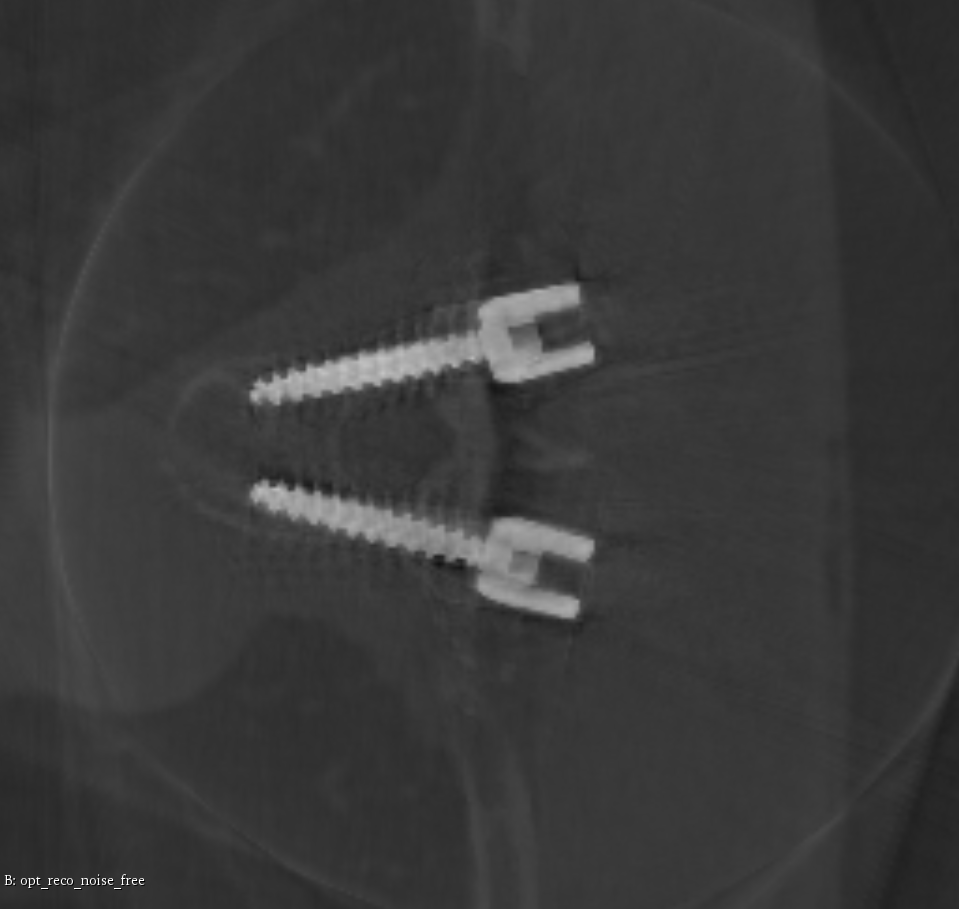}
\caption{}
\end{subfigure}
\begin{subfigure}[t]{0.24\textwidth}
\centering
\includegraphics[trim= 5 11 0 15, clip, width=\textwidth]{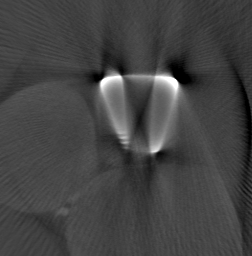}
\caption{}
\end{subfigure}
\begin{subfigure}[t]{0.24\textwidth}
\centering
\includegraphics[trim= 5 6 0 15, clip, width=\textwidth]{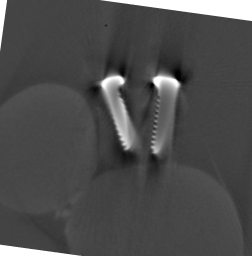}
\caption{}
\end{subfigure}
\caption{Representative axial slices through reconstructed volumes for the synthetic test set at vertebral level T6 in (a,b), and for the real data set in (c,d). Slices (a,c) show images using a straight-forward short-scan, while slices (b,d) show results of a recommended task-aware trajectory.}
\label{fig:synthetic_results}
\end{figure}

\paragraph{Qualitative Synthetic Data Experiments:} From a clinical perspective the quality of the reconstructions is most interesting. For the synthetic test data, representative reconstructions from a short-scan and a task-aware trajectory are shown in \Fref{fig:synthetic_results}(a) and (b), respectively. Both volumes were reconstructed using iterative conjugated gradients least-squares (ASTRA Toolbox) from noise free projections. It is apparent that, for the proposed task-aware trajectory, the screw is more homogeneous, exhibits less cupping artifact, and metal artifacts (bright and dark streaks) are reduced enabling better assessment of bony anatomy in close proximity to the implant.

\paragraph{Real Data Experiments:} Real data experiments are performed on a physical phantom consisting of two ballistic gel cylinders mounted on a wooden beam with two iron screws, abstractly mimicking the human chest. A C-arm system (Siemens Arcadis Orbic 3D) was used to acquire five CBCT short-scans, performed with inclinations ranging from large negative $(\sim-30\deg)$ to large positive values $(\sim30\deg)$. Corresponding reconstructions obtained with filtered back-projection~\cite{feldkamp1984practical} were also obtained from the system.

\begin{figure}[tb]
    \centering
    \includegraphics[trim=0 0 0 0, clip, width=0.4\linewidth]{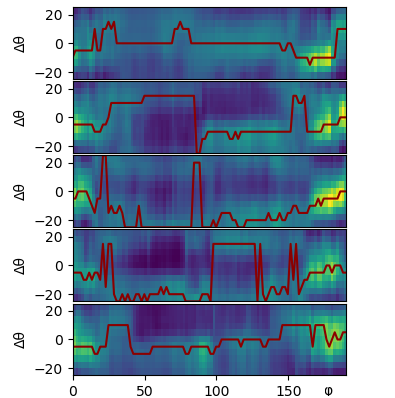}
    \caption{Detectability maps predicted from real X-ray projection images of a simple human chest model. From top to bottom, inclination of the scan is decreasing; the midline of each map corresponds to the scan axis. The red curve indicates the optimal trajectories predicted from the map. Figure best viewed zoomed in.}
    \label{fig:real_detectability_maps}
\end{figure}

The detectabilities of next possible views are predicted with the ConvNet trained on our analytic phantom dataset. The resulting predictions are shown in \Fref{fig:real_detectability_maps}. As the available C-arm geometry currently limits the reconstruction to circular trajectories, we compare the quality of the best circular scan with the conventional circular scan. The best circular scan is determined by accumulating the predicted detectability for $\Delta\Theta=0$, which closely models the overall task. Via this approach, the scan with the highest positive tilt is selected, yielding a $19.0\%$ increase in predicted detectability compared to the conventional scan. 

In \Fref{fig:real_detectability_maps}, we highlight the highest predicted detectability at any given time, which corresponds to the unregularized servoing commands that would be sent to the C-arm. Curves close to the centerline indicate little C-arm adjustments, while curves far from it imply our agent trying to drive the C-arm out of the central plane. We observe large out-of-plane angle commands for scans with low absolute tilt (conventional), and minimal deviation for scans with high tilt, indicating a close to optimal trajectory. This behavior is well interpretable and fits our intuition: The algorithm tries to prevent images with screw overlap, thus reducing metal artifact in the reconstruction. Slices in the screw-plane, reconstructed from the conventional and recommended trajectory are shown in \Fref{fig:synthetic_results}(c) and (d), respectively. The reconstruction from the high-tilt trajectory recommended by our system exhibits a notable reduction of metal artifacts and noise, and reveals the screw thread that is completely invisible in the conventional case. We anticipate overall image quality improvements when using C-arm systems with flat-panel detectors and more brilliant X-ray sources.

\section{Conclusion}
We present the first step towards task-aware CBCT imaging protocols. Our approach enables scene-specific source trajectories in clinical settings, where only little prior information is available. On both synthetic and real CBCT scans,
in simulations with human anatomy as well as on real X-ray images of a phantom, we demonstrate that task-aware image acquisition is a promising avenue for prospectively improving image quality in CBCT reconstruction. The proposed learning-based method recommends viewing angles onto anatomy and can be combined with any (iterative) reconstruction or metal artifact reduction algorithm. In future work, we will test our system on cadaveric specimens, pedicle screws on multiple levels, and more complex tasks including soft tissue imaging.

\vspace{5pt}
\noindent \textbf{Acknowledgemens:} \ifanonymized No acknowledgements given at this time. \else We gratefully acknowledge support of the NVIDIA Corporation for donating GPUs, and Gerhard Kleinzig and Sebastian Vogt from SIEMENS for making an ARCADIS Orbic 3D available. JNZ was supported by a DAAD FITweltweit fellowship. \fi

%
%
%
%
\bibliographystyle{splncs04}

\end{document}